# ANÁLISE DA DECIMAÇÃO NO PROCESSAMENTO DE SINAIS DE CORRENTE PARA DETECÇÃO DE BARRAS QUEBRADAS EM MOTORES DE INDUÇÃO

Josemar S. Moreira*, Paulo C. M. Lamim Filho*, Lane M. R. Baccarini*, Erivelton G. Nepomuceno**, Priscila F. S. Guedes**

*LAMET - Laboratório de Máquinas Elétricas e Transformadores,*
*\** *GCOM - Grupo de Controle e Modelagem,*
*Departamento de Engenharia Elétrica, Universidade Federal de São João del-Rei*
*Rua Frei Orlando, 170 – Centro, São João del-Rei, Minas Gerais, CEP: 36307-352*
*E-mails:* `js_moreira99@yahoo.com.br`

**Abstract** — This paper presents a study on the reduction of the sampling frequency of the current signals of an induction motor, the reductions are performed by means time-decimation technique for digital signal processing. We have used the Fast Fourier Transform to obtain the fault signal spectrum of broken bars. The results have shown how the decimation technique significantly reduces the number of operations and the time required to calculate the Fast Fourier Transform without loss of information. This approach provides a better performance of embedded systems for fault diagnosis based on characteristic of amplitude signal modulation.

**Keywords** — Decimation, signal processing, frequency spectrum, broken bars, induction motors.

**Resumo** — Este artigo traz um estudo sobre a redução da frequência de amostragem dos sinais de corrente de um motor de indução, sendo as reduções realizadas pela técnica de decimação no tempo para o processamento digital de sinais. Utilizamos a Transformada Rápida de *Fourier* para obter o espectro do sinal de falha de barras quebradas. Os resultados mostram como a técnica de decimação reduz significativamente o número de operações e o tempo necessário para o cálculo da Transformada Rápida de *Fourier* sem perda de informação. Esta abordagem pode proporcionar um melhor desempenho de sistemas embarcados para o diagnóstico das falhas baseadas na característica de modulação de sinais em amplitude.

**Palavras-chave** – Decimação, processamento de sinais, espectro de frequência, barras quebradas, motores de indução.

## 1 Introdução

As máquinas elétricas assíncronas de rotor gaiola, são largamente utilizadas para o acionamento de cargas mecânicas uma vez que podem trabalhar em ambientes agressivos pois não possuem anéis no rotor, tem baixo custo de aquisição e de manutenção em comparação às demais máquinas elétricas. Assim, os motores de indução do tipo gaiola de esquilo representam aproximadamente 85% dos motores instalados nas indústrias (Abd-el-Malek, 2017). Devido a suas aplicações, os motores estão sujeitos a estresses térmicos, mecânicos e elétricos que podem ocasionar em falhas durante sua operação. Nos últimos 30 anos, pesquisadores e empresas têm desenvolvido várias técnicas que podem ser usadas para detecção de falhas em motores elétricos. O objetivo principal é ter uma técnica confiável, mas também de fácil aplicação industrial, de modo a não colocar o operador em contato direto com a máquina monitorada. Um método bastante utilizado é a Análise de Padrões da Corrente do Motor (MCSA), (Puche-Panadero, 2012).

A técnica MCSA, baseia-se na análise espectral da corrente do estator, utilizando como ferramenta matemática a Transformada Rápida de *Fourier* (FFT) que possibilita a identificação de padrões que correspondem a assimetrias ou falhas no rotor de motores de indução. Este método de detecção de falhas é amplamente utilizado devido à sua característica não invasiva que possibilita a aquisição de sinais sem a necessidade de acesso a parte interna do motor ou parada do processo produtivo (Fang, 2006).

Porém, a MCSA tem suas limitações. Em Gyftakis (2014) é relatada a dificuldade de diagnosticar falhas de barras quebradas quando o motor opera com baixas cargas, isto ocorre devido a assinatura de falha ser de baixa frequência. A mesma dificuldade é relatada por Sapena-Bano (2015) e Puche-Panadero (2009), que utilizam um longo tempo de amostragem de 100 segundos para conseguir uma resolução em frequência de 0,001 Hz. Isto implica na necessidade de aumento do espaço de memória e do tempo do processamento dos sinais, exigindo grande esforço computacional.

Uma saída para este problema é o uso da decimação, que consiste na redução da taxa de amostragem de um sinal por um fator qualquer (Crochiere, 1981). Crochiere e Rabiner (1975) realizaram estudos quantitativos do custo computacional de sistemas diante da redução ou aumento da taxa de amostragem. Ding (2011) utilizou um filtro de decimação digital para o processamento de sinais em tempo real de veículos micro aéreos, obtendo resultados satisfatórios em sistemas de controle. Tan (2014) e Robles (2015) também aplicaram esta técnica a filtros digitais. Chakkor (2015) utiliza a decimação em um algoritmo para detecção de falhas de turbinas eólicas. Os resultados do estudo demonstraram o desempenho do algoritmo permitindo a identificação de harmônicos de alta resolução com tempo mínimo de computação e menor custo de memória.

Apesar das aplicações da decimação apresentarem bons resultados em termos de velocidade e consumo de memória, existem poucos trabalhos relacionados ao processamento de sinais na detecção de falhas em motores de indução com este objetivo. Neste sentido, a proposta desta pesquisa consiste em reduzir a frequência de amostragem de um sinal de corrente, afim de melhorar o processamento de sinais na detecção de falhas de barras quebradas em motores e indução. Com o processo de decimação, espera-se diminuir ao máximo possível a taxa de amostragem dos sinais de corrente, sem comprometer a informação do sinal de falha, visando reduzir o esforço computacional e o espaço ocupado em memória.

## 2 Fundamentação Teórica

### 2.1 Falha de barras quebradas e/ou trincadas do Motor

A ocorrência de barras quebradas em motores de indução provoca modulações em amplitude nos sinais de corrente do estator. Estas modulações possuem frequências características de falha que podem ser obtidas através de um processo de demodulação de sinais (Bonnett, 2008).

Tais frequências características de falha, são representadas pela Equação 1, na qual as constantes *s* e *f* referem-se, respectivamente, ao escorregamento e a frequência de alimentação, sendo *k* um número inteiro positivo.

$$f_{br} = (1 \pm 2sk)f \quad (1)$$

### 2.2 Transformada de Hilbert

A Transformada de *Hilbert* (HT) é uma técnica de análise de sinais bem conhecida e amplamente usada em diferentes campos científicos, como transmissão de sinais, processamento de dados geofísicos, e demodulação de sinais em motores de indução (Puche-Panadero, 2009).

A transformada de Hilbert de um sinal modulado qualquer representado por *x(t)*, é definida pela Equação 2.

$$H[x(t)] = \frac{1}{\pi}\int_{-\infty}^{+\infty}\frac{x(t)}{t-\tau}d\tau \quad (2)$$

Pode-se criar um sinal analítico *z(t)* com parte real e imaginária, sendo a parte real o próprio sinal e a parte imaginária a Transformada de *Hilbert* deste sinal. Na forma polar o sinal analítico é definido pela Equação 3.

$$z(t) = A(t)e^{j\theta(t)} \quad (3)$$

Sendo que *A(t)* e *θ(t)* representam a amplitude e a fase instantânea do sinal analítico, respectivamente.

O envelope ou sequência de tempo modulada de *z(t)* é definido como o módulo do sinal analítico.

### 2.3 Decimação

A técnica de decimação consiste em reduzir o número de amostras originais de um sinal por um fator inteiro *p*. Tal processo pode ser feito tomando uma amostra e eliminando as demais amostras *p-1*, ou preferencialmente calculando a média da amostra *p*, pois, desta forma reduz-se o ruído de quantificação. Nesse sentido, a decimação funciona como um filtro passa baixa, eliminando a banda de alta frequência do sinal original. Ao longo da frequência de amostragem reduzida e largura de banda, a sequência de dados também é diminuída em *p* vezes (De Jesus, 2017).

O exemplo a seguir mostra a aplicação da técnica em um dado sinal no domínio do tempo discretizado a uma frequência de amostragem de 2500 Hz, como ilustrado na Figura 1. Ao aplicar um fator de decimação *p=4* obtêm-se o gráfico da Figura 2.

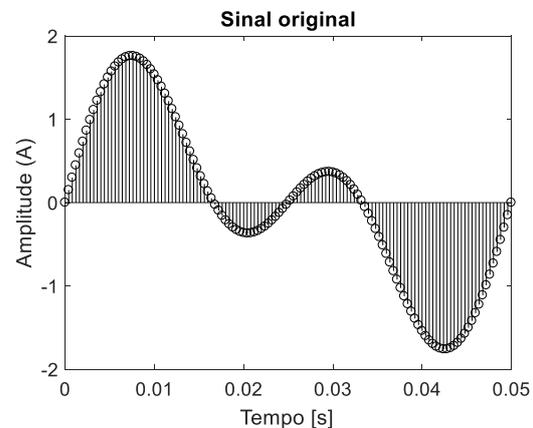

Figura 1 - Sinal amostrado a 2500 Hz.

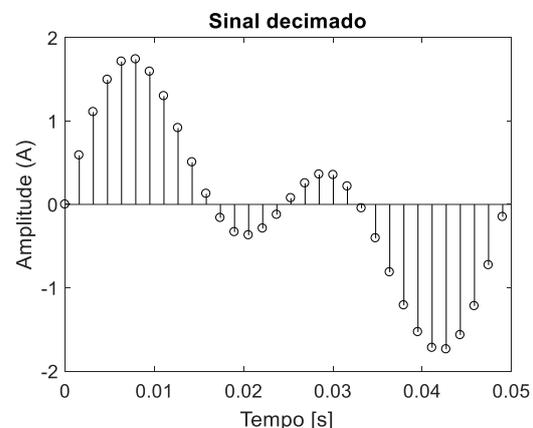

Figura 2 – Sinal decimado por um fator 4.

Nota-se que o sinal possui mesmo formato, porém um número menor de pontos, como se estivesse sido amostrado a uma frequência de 625 Hz. Desta forma, com um número menor de pontos é possível representar o mesmo sinal mantendo suas informações de magnitude e forma de onda.

## 3 Metodologia

A metodologia proposta para a redução da taxa de amostragem do sinal de corrente consiste nas seguintes etapas:

1. Aplicar um filtro passa-faixa ao sinal de corrente amostrado com as frequências de corte $f_1 = 40$ Hz e $f_2 = 70$ Hz, separando assim a faixa de interesse para estudo.
2. Decimar o sinal de corrente filtrado por um fator $p_i = 2^n$, sendo $n$ um número inteiro inicialmente igual a 1.
3. Aplicar a Transformada de *Hilbert* ao sinal decimado, obtendo o envelope (sinal demodulado) e retirar a componente DC.
4. Aplicar a FFT ao sinal demodulado, obtendo o espectro do sinal de falha.
5. Observar o espectro do sinal. Caso possua o mesmo valor encontrado para o sinal original (sem decimação), retornar ao passo 2 e decimar por um fator $p_i = 2^{n+1}$. Caso o valor obtido seja diferente, encerra-se o processo.

Para realização das etapas descritas foi utilizado o *software* Matlab R2013. A aquisição dos sinais de corrente de estator do motor de indução foi realizada em uma bancada (Figura 3) composta por um motor de indução trifásico WEG 3 CV, 220 V, 60 Hz, 4 polos e velocidade nominal de 1735 rpm. O motor em questão foi acoplado a um gerador CC ligado a um banco de resistores que em conjunto, funcionam como um sistema de carga para o motor de indução. Entre o motor de indução e o gerador foi inserido um transdutor de torque, ligado a um torquímetro digital para medir o nível de torque e garantir uma aquisição de dados padronizada. Na Figura 4 tem-se o rotor com um furo sobre uma barra, simulando a condição de falha.

Os sinais de corrente foram coletados por meio de um alicate amperímetro (A622 AC/DC - 100 Hz Tectronix), ligado a uma placa de aquisição (NI PCI-4461 da *National Instruments*. A placa possui duas entradas analógicas simultâneas de 24-bits (taxa de amostragem máxima de 80 kHz) e duas saídas analógicas simultâneas de 24-bits (taxa de amostragem de 204,8 kS/s) com faixa de entrada de ±316 mV até 42,4 V. Além disso, ela possui filtro *anti-aliasing* até 92 kHz.

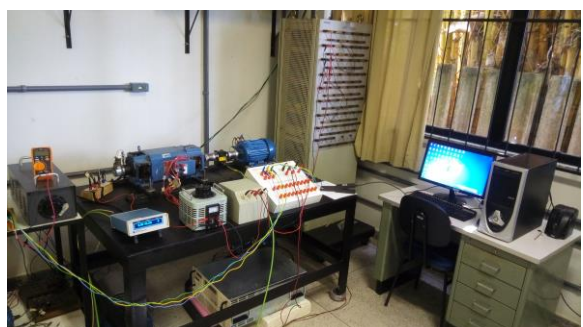

Figura 3 - Bancada de testes experimentais.

Os sinais de corrente foram coletados a uma frequência de amostragem de 5,12 kHz, durante 102,4 segundos, gerando um total de 524288 amostras.

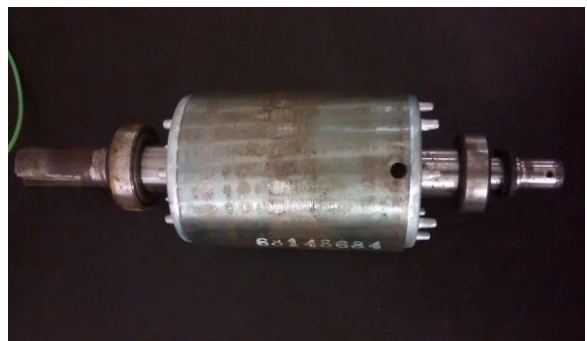

Figura 4 - Rotor com uma barra quebrada.

## 4 Resultados

Foram analisados 10 sinais de corrente coletados em uma das fases de alimentação do motor operando à 50%, 80% e 100% de carga. Na Tabela 1 tem-se o número de amostras e as respectivas frequências dos sinais para cada fator de decimação aplicado.

Tabela 1. Número de amostras e frequências dos sinais decimados.

| Fator de decimação | Número de amostras | Frequência de amostragem (Hz) |
|---|---|---|
| 1 | 524288 | 5120 |
| 2 | 262144 | 2560 |
| 4 | 131072 | 1280 |
| 8 | 65536 | 640 |
| 16 | 32768 | 320 |

Sabendo-se que os sinais de falha de barras quebradas estão localizados nas baixas frequências, teoricamente seria possível decimar o sinal até o fator $p_i = 32$, com a frequência artificial de amostragem de 160 Hz, respeitando o teorema de *Nyquist*, sem perda de informação em termos de frequência e magnitude do sinal. Porém, tal fator ao ser aplicado provocou o deslocamento da frequência característica de falha além de apresentar um erro médio de 2,8% no espectro de corrente. Sendo aceitável para análise de falhas um erro médio de 1%, considerou-se a decimação até o fator $p_i = 16$, que caracterizou um erro médio de 0,96% como visto na Figura 5.

Nas Figuras 6, 7 e 8 tem-se os espectros de corrente dos sinais de falha para o motor operando a 50%, 80% e 100% de carga, respectivamente. Obtidos com o sinal original coletado a 5,12 kHz. Já nas Figuras 9, 10 e 11 tem-se os espectros de corrente do sinal de falha decimado por um fator $p_i = 16$. Nota-se apenas uma pequena variação da magnitude do sinal original para o sinal decimado. Entretanto, ao aplicar o fator $pi = 32$, toda a informação é perdida, tanto em magnitude quanto em frequência como visto na Figura 12.

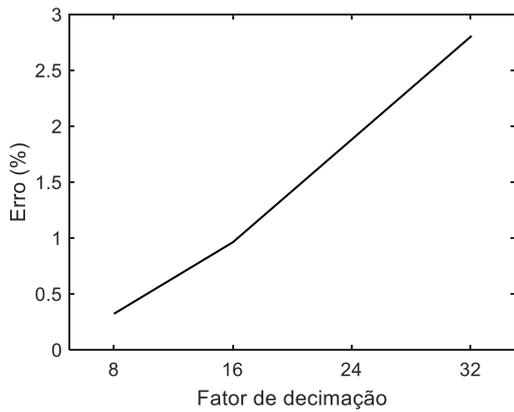

Figura 5 - Erro médio relacionado ao fator de decimação.

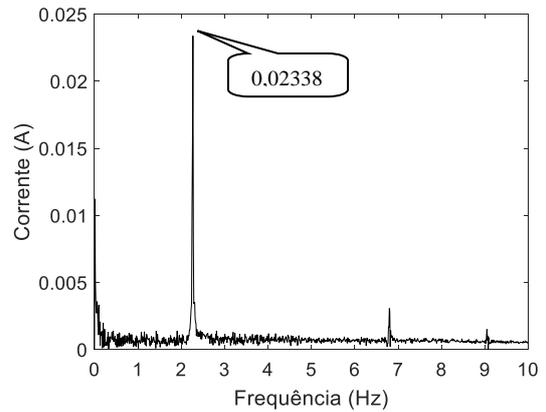

Figura 9 - Espectro do sinal decimado a *pi = 16* e motor a 50% de carga.

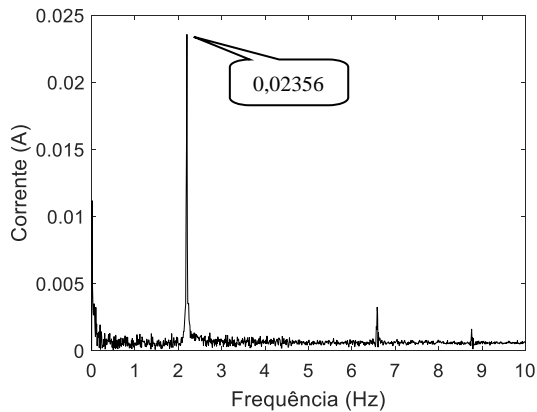

Figura 6 - Espectro do sinal original a 50% de carga.

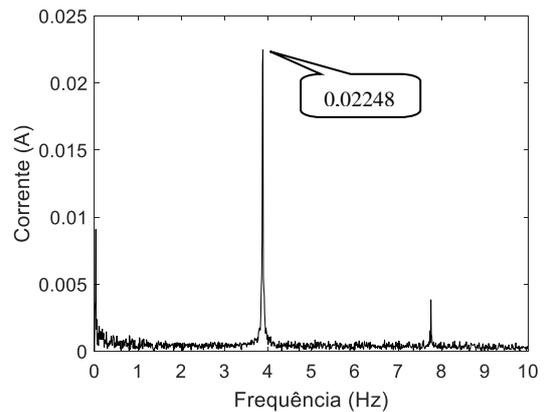

Figura 10 - Espectro do sinal decimado a *pi = 16* e motor a 80% de carga.

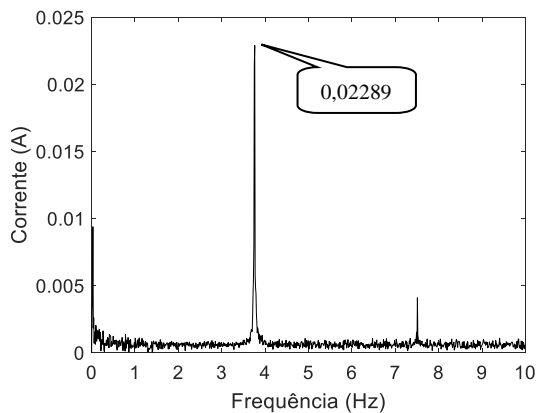

Figura 7 - Espectro do sinal original a 80% de carga.

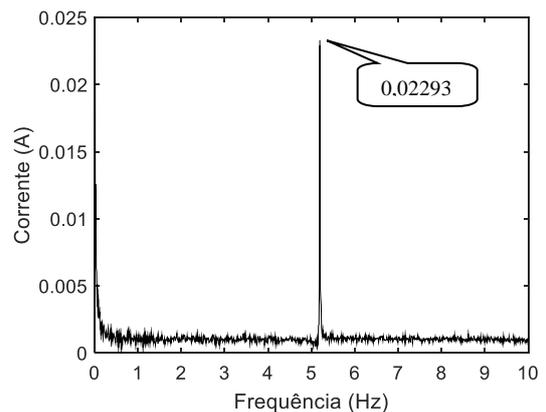

Figura 11 - Espectro do sinal decimação a $p_i = 16$ e motor a 100% de carga.

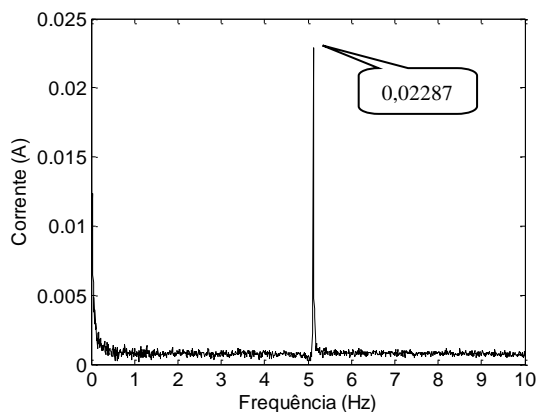

Figura 8 - Espectro do sinal original a 100% de carga.

O algoritmo proposto por Colley e Tukey (1965), tradicionalmente conhecido como Transformada Rápida de *Fourier* (FFT) para o cálculo computacional da Transformada Discreta de *Fourier* (DFT), reduz drasticamente o número de operações da ordem de $N^2$ para $Nlog_2N$ sendo $N$ uma potência de 2 (Duhamel, 1990). Conforme Lathi (2006) são necessárias $Nlog_2N$ adições complexas e $(N/2)log_2N$ multiplicações complexas para o cálculo da DFT via algoritmo FFT.

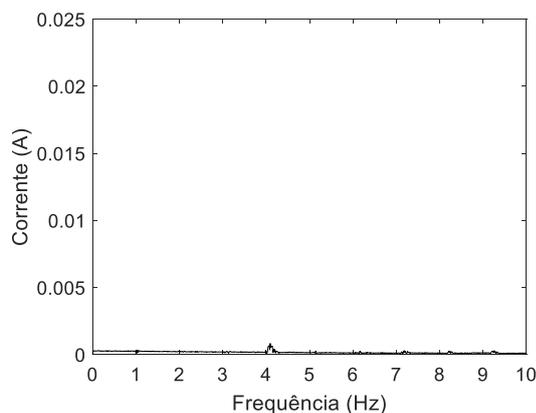
Figura 12 - Espectro de corrente do sinal com fator pi = 32 e motor a 100% de carga.

Neste contexto, a redução do esforço computacional com o algoritmo de identificação de falhas proposto, pode ser percebida na Tabela 2 que relaciona o número de operações necessárias para o cálculo da FFT referente a cada fator de decimação aplicado. O máximo fator de decimação utilizado reduziu em 20 vezes o número de operações necessárias para o cálculo da FFT.

Tabela 2. Número de operações necessárias para o cálculo da FFT.

| Fator de decimação | Número de multiplicações | Número de adições | Total de operações |
|---|---|---|---|
| 1 | 4980736 | 9961472 | 14942208 |
| 2 | 2359296 | 4718592 | 7077888 |
| 4 | 1114112 | 2228224 | 3342336 |
| 8 | 524288 | 1048576 | 1572864 |
| 16 | 245760 | 491520 | 737280 |

A redução do número de operações, proporcionada pelo algoritmo, sugere uma redução também no tempo de processamento para o cálculo da FFT. A fim de mensurar esta redução, foram realizadas 30 simulações com cada um dos 10 sinais de falha aqui estudados, obtendo-se um tempo médio para cada fator de decimação aplicado como mostrado na Tabela 3.

Tabela 3. Tempo médio de processamento para o cálculo da FFT.

| Fator de decimação | Número de amostras | Tempo médio (ms) |
|---|---|---|
| 1 | 524288 | 28,55 |
| 2 | 262144 | 13,53 |
| 4 | 131072 | 6,52 |
| 8 | 65536 | 2,58 |
| 16 | 32768 | 1,35 |

## 5 Conclusão

Neste trabalho é apresentada uma análise do uso da decimação no tempo para o processamento de sinais de falha em motores de indução. Os resultados mostram como é possível reduzir significativamente o número de amostras de um sinal, sem perder as informações de falha nele contidas. Com a máxima redução do número de amostras, o número de operações necessárias para o cálculo da FFT foi 20 vezes menor e o tempo médio para o cálculo da FFT foi reduzido de 28,55 para 1,35 ms.

A técnica de decimação no tempo pode proporcionar um melhor desempenho de sistemas embarcados para o diagnóstico de falhas que tenham como característica a modulação de sinais em amplitude.

Apesar deste trabalho abordar um estudo sobre detecção de falhas, o uso desta técnica não se restringe apenas a esta área. Como encontrado na literatura, a modulação em amplitude é usada em diversas áreas como sistemas de telecomunicações e teleproteção de sistemas elétricos de potência, onde a aplicação da decimação no tempo para o processamento digital de sinais pode ser aplicada, visando um melhor desempenho computacional dos sistemas.